\documentclass[10pt,conference]{IEEEtran}
\IEEEoverridecommandlockouts

\usepackage{cite}
\usepackage{amsmath,amssymb,amsfonts}
\usepackage{algorithmic}
\usepackage{graphicx}
\usepackage{textcomp}
\usepackage{xcolor}
\usepackage[hidelinks]{hyperref}
\def\BibTeX{{\rm B\kern-.05em{\sc i\kern-.025em b}\kern-.08em
    T\kern-.1667em\lower.7ex\hbox{E}\kern-.125emX}}

\begin{document}
\begin{sloppy}

\title{Bot-Driven Development: From Simple Automation to Autonomous Software Development Bots}

\author{\IEEEauthorblockN{Christoph Treude}
\IEEEauthorblockA{\textit{School of Computing and Information Systems} \\
\textit{Singapore Management University}\\
Singapore  \\
ctreude@smu.edu.sg}
\and
\IEEEauthorblockN{Christopher M. Poskitt}
\IEEEauthorblockA{\textit{School of Computing and Information Systems} \\
\textit{Singapore Management University}\\
Singapore  \\
cposkitt@smu.edu.sg}
}

\maketitle

\begin{abstract}
As software development increasingly adopts automation, bot-driven development (BotDD) represents a transformative shift where bots assume proactive roles in coding, testing, and project management. In bot-driven development, bots go beyond support tasks, actively driving development workflows by making autonomous decisions, performing independent assessments, and managing code quality and dependencies. This paper explores how bot-driven development impacts traditional development roles, particularly in redefining driver-navigator dynamics, and aligns with DevOps goals for faster feedback, continuous learning, and efficiency. We propose a research agenda addressing challenges in bot-driven development, including skill development for developers, human-bot trust dynamics, optimal interruption frequency, and ethical considerations. Through empirical studies and prototype systems, our aim is to define best practices and governance structures for integrating bot-driven development into modern software engineering.
\end{abstract}


\section{Who is the Driver?}

In traditional pair programming, collaboration revolves around two well-defined roles: the driver and the navigator~\cite{williams2001integrating}. The driver actively writes code, focusing on the technical details and syntax, while the navigator maintains a broader view, offering feedback on design, strategy, and potential problems. This complementary dynamic allows both team members to contribute unique perspectives: the driver handling the fine details, and the navigator maintaining a strategic overview.

However, with tools like GitHub Copilot---tagged as ``Your AI pair programmer''~\cite{bird2022taking}---this clear division begins to blur. On the one hand, Copilot could be seen as a ``navigator'', reacting to the developer’s input and providing suggestions that align with the immediate coding context. But unlike a human navigator, Copilot does not have an understanding of the project’s overarching goals, design principles, or strategic objectives. It operates reactively, offering snippets that fit syntactically or structurally, without questioning intent or suggesting alternative approaches based on higher-level considerations.

Labeling Copilot as a navigator is tempting, but it lacks the true qualities of a navigator. A traditional navigator actively monitors the driver’s work, contributes ideas to shape the broader direction of the task, and might even stop progress to suggest a more effective approach~\cite{chong2007social}. However, Copilot does not interrupt or interject with strategic recommendations; it functions silently, offering suggestions only when prompted by the context of the code being written. The developer ultimately decides whether to accept, reject, or modify these suggestions, effectively making them both a driver and a navigator.

Conversely, it is not accurate to consider Copilot the ``driver'' and the developer the ``navigator''. While Copilot writes code snippets, it does so only in response to cues from the human developer; it cannot initiate, plan, or consider the implications of its suggestions. The developer actively guides the direction of the code, prompts Copilot through their own writing, and is responsible for aligning its contributions with the larger objectives. Therefore, while Copilot contributes to the coding process, it lacks the autonomy, awareness, and decision-making capabilities to function as a true driver~\cite{plonka2011collaboration}.

In this interaction, the developer is both driver and navigator, relying on Copilot for small-scale, context-specific suggestions without the broader strategic contribution that a human navigator might provide. This blended role creates a unique form of co-driven development, where the human still directs the task’s path but with continuous quiet help from AI~\cite{wang2020human}.

Beyond Copilot, the landscape of bots in software development is diverse, ranging from tools that offer syntax support and code generation to those that streamline version control, automate testing, and assist with deployment~\cite{wessel2018power}. For example, Renovate\footnote{\url{https://www.mend.io/renovate/}} is a bot that manages dependency updates, automatically opening pull requests when libraries in a project are out of date. Common examples include linting bots that enforce coding standards, dependency management bots that monitor and update third-party libraries, and issue triaging bots that help manage and prioritize reported issues in repositories. Each of these bots takes on specialized tasks, contributing to the overall development process but typically in discrete, well-defined ways~\cite{santhanam2022bots}.

Most bots in today’s ecosystem perform their tasks independently of the core development activity, often activating at specific stages or as scheduled jobs. Their interactions with developers are largely non-intrusive; they either automate routine maintenance or perform quality assurance tasks that occur after the primary code has been written~\cite{wessel2022bots}. For example, ESLint\footnote{\url{https://eslint.org/}} is a popular linting bot that enforces JavaScript code style and best practices, automatically flagging issues for developers. This range of functionalities has led to a bot-assisted development environment where human developers remain in charge, and bots perform supportive ancillary roles.

One prominent category of these bots is Continuous Integration (CI) bots, which generally represent the final step in the development process. CI bots automate testing, code quality checks, and integration validations, functioning as a last line of defense before the code is merged or deployed. For example, GitHub Actions\footnote{\url{https://github.com/features/actions}} can be configured as a CI bot that runs tests on code changes and verifies that they meet quality standards before merging~\cite{wessel2023github}. Unlike Copilot, CI bots are not active collaborators in the development flow itself; they only engage once the human developer has completed the main development work. In this sense, CI bots act more as quality gatekeepers than co-drivers. They enforce standards and identify potential problems, but they do not contribute to the creative aspects of problem solving in development.

As we move toward bot-driven development, these distinctions challenge us to rethink driver and navigator roles. In a future where more intelligent, autonomous bots participate actively in all stages of development, we may see an evolution from passive to interactive bot roles. Developers could work with bots that dynamically engage at both the micro- and macro-levels, providing input that extends beyond syntax suggestions to design insights, performance recommendations, and architectural decisions. This shift will redefine both agency and collaboration, positioning bots not just as tools but as active partners in development.

\section{Bot-Driven Development}

In bot-driven development (BotDD), bots play an active role, shaping the development process in real time. Unlike the current landscape, where bots mostly provide passive assistance, bot-driven development positions them as proactive agents in the software creation workflow. These bots not only automate routine tasks, but actively manage, assess, and drive progress. This aligns closely with the DevOps goals of faster flow, fast feedback, and continuous learning~\cite{fitzgerald2014continuous,kim2021devops}. By embedding immediate feedback directly into the development cycle, bot-driven development allows issues to be identified and addressed before they are passed further down the stream. Rather than simply waiting for human direction, bots in a bot-driven development setting can continuously monitor code, review standards, and make autonomous decisions on prioritizing fixes or improvements, accelerating feedback loops and reducing bottlenecks.

In a bot-driven development environment, bots are envisioned as running continuously in the background, assessing the state of code, dependencies, and potential risks, and even assigning urgency levels to tasks. For example, imagine a security bot that detects a vulnerability in a newly merged code update. Instead of merely notifying the developer, this bot could independently evaluate the severity of the issue, consider its potential impact on other systems, and prioritize it for immediate resolution or schedule it for later if it is less critical. This kind of autonomous prioritization helps teams stay focused on critical issues while still addressing smaller improvements over time. Another example might involve a bot that monitors library dependencies. If a dependency becomes outdated or potentially insecure, the bot could evaluate whether an update aligns with the current project goals and then automatically apply, test, and deploy the update if the assessment is favorable. This autonomy in managing critical aspects of the codebase brings us closer to a scenario where bots are not just tools but operational team members actively engaged in ensuring high-quality, safe, and maintainable software.

What could this look like in practice? Here are some emerging dynamics that we can anticipate:

First, there is a potential for multi-agent bot collaboration~\cite{he2024llm}. Rather than a single bot handling all the responsibilities, different specialized bots could collaborate as a team. Each bot might focus on a specific aspect of development: one continuously optimizes the code for performance, another tracking dependencies, and a third monitoring code for security issues. Imagine a scenario where a security-focused bot flags a potential vulnerability, prompting the dependency bot to assess if a new library version can address it, and a performance bot to determine if this update has performance impacts. The bots then coordinate to make a collective decision on whether to proceed with an update, with the dependency bot initiating it if the consensus is reached. This collaborative network of bots could act as an intelligent, adaptive ecosystem, bringing in precise and timely interventions without requiring human intervention for every decision.

This shift also reimagines the role of the developer as a bot moderator. In bot-driven development, developers would spend less time on manual coding and debugging and more time orchestrating and guiding the collective contributions of multiple bots~\cite{shihab2022present}. They might set high-level goals, monitor how bots interpret these goals, and intervene to correct or adjust the course when necessary. For example, a developer might monitor a real-time dashboard that visualizes the activities of different bots and how their actions align with project milestones. If a bug detection bot starts to focus a lot on minor cosmetic issues rather than critical bugs, the developer could adjust the bot’s parameters to prioritize high-severity errors. In this way, the developer is less involved in day-to-day coding and more engaged in a high-level management role, fine-tuning, and supervising bot interactions.

Looking further ahead, bot-driven development may eventually pave the way toward autonomous development~\cite{rasheed2024codepori}. In this vision, bots evolve to handle entire coding tasks independently, from writing and testing to integrating and deploying code changes. For example, a team could begin a project by defining a set of specifications, and from there a group of bots would autonomously develop features, run tests, and even optimize performance based on usage data. Bots could check and refactor each other’s code and even suggest new features based on analytics, learning over time to better align with user needs. Here, developers would move almost entirely into an oversight role, periodically checking that the direction of the system aligns with the project’s goals and stepping in only when complex, high-level adjustments are required.

This trajectory toward autonomy, multi-agent collaboration, and bot moderation marks a profound shift in software development. In bot-driven development, developers are no longer just creators of code but facilitators, orchestrators, and vision-setters. By transforming the development process from human-driven to bot-driven, bot-driven development has the potential to not only improve speed and accuracy but also create a sustainable, adaptive workflow where bots become reliable, independent contributors in the ongoing evolution of software.

\section{Research Agenda}

The shift to bot-driven development introduces a series of research challenges that need to be addressed to fully leverage the potential of autonomous and collaborative bots in software development. In the following, we outline key challenges, research questions, and possible approaches to explore these questions.

\subsection{Implications for Skill Development in Bot-Driven Teams}

As bots take on more active roles in development, the skills required from developers will shift~\cite{ma2024teach}. Developers will need new competencies, such as configuring, debugging, and monitoring bots in real-time. Understanding how these skills evolve is crucial to preparing developers for bot-driven environments.

\textbf{Research Questions:} What skills are required for developers working in bot-driven teams? How can developers best moderate and orchestrate multi-bot systems? What new competencies (e.g., bot configuration, bot debugging, workflow monitoring) will developers need, and how should educational programs evolve to support these competencies?

Surveys and interviews with developers working with advanced bot-driven systems could help identify emerging skill requirements. Controlled experiments in educational settings where students are exposed to bot-driven workflows can provide insight into how skill development progresses. Furthermore, longitudinal studies could track skill acquisition and adjustment as developers gain experience with bot-driven environments, highlighting areas that need targeted training or additional support.

\subsection{Human-Bot Trust and Collaboration Dynamics}

Establishing trust in bots is essential for effective collaboration, especially as bots become more autonomous in making decisions~\cite{erlenhov2020empirical}. Factors such as transparency, predictability, and explainability of bot actions play a critical role in building and maintaining this trust.

\textbf{Research Questions:} What factors influence developer trust in autonomous bots within bot-driven development? How can bot behavior, transparency, and explainability be optimized to build and maintain trust? How does the predictability or reliability of bot-driven decisions affect human-bot collaboration?

To address these questions, experiments can be designed that vary bot transparency levels, confidence indicators, and error rates, observing how these factors impact developer trust and performance. Ethnographic studies could provide deeper insights by observing developer interactions with bots over time in real-world settings. Cognitive walk-throughs and usability testing can assess how interface design influences trust and collaboration dynamics.

\subsection{Tuning Frequency of Interruptions and Confidence of Bots}

In bot-driven development, bots need to provide timely assistance without overwhelming developers. Finding the right balance in the frequency of bot interruptions and confidence indicators is key to ensuring productive workflows and minimizing disruption~\cite{parnin2011resumption}.

\textbf{Research Questions:} How can bots be tuned to provide the right amount of feedback or interruptions to optimize productivity? What levels of confidence should bots demonstrate to effectively support developer decision-making without overwhelming or disrupting the workflow? What metrics can be used to gauge appropriate interruption levels?

User studies that vary the frequency of interruption and confidence indicators between bot setups could reveal the most effective configurations. Real-time metrics, such as developer engagement and task flow, could be analyzed to correlate bot behavior with productivity and satisfaction. Machine learning models trained on these metrics could allow bots to adjust the frequency of interruption adaptively, customizing interactions based on developer preferences.

\subsection{Workflow Integration and Interfaces}

Seamlessly integrating bots into development workflows is critical to maximizing their utility. Effective interfaces are needed to support interaction with multiple bots, facilitate smooth oversight, and minimize cognitive load for developers~\cite{yuan2024maxprototyper}.

\textbf{Research Questions:} How should bots be integrated into existing software development workflows? What are the optimal interface designs for bot-driven systems to support seamless, transparent interactions? How can multi-bot collaboration interfaces be designed to facilitate human oversight and intervention without cluttering the developer experience?

Participatory design sessions with developers could help explore the preferences of the interface and workflow, offering practical insights into effective bot integration. Usability testing with prototypes would allow researchers to assess the impact of different interaction models, such as dashboards, embedded notifications, and multi-bot status displays. Real-world case studies in development environments could provide feedback on interface effectiveness and suggest further adjustments to improve the developer experience.

\subsection{Customization and Project Constraints}

Bots need to operate within specific project constraints, such as coding standards, security requirements, and performance goals. Allowing customization of bot behaviors for different project needs is crucial to ensure that bots act in line with project-specific boundaries~\cite{hidaka2019design}.

\textbf{Research Questions:} How can bot behaviors be customized to align with unique project constraints, such as security requirements, performance goals, or coding standards? What customization options do developers need to ensure that bots act within project-specific boundaries? How can bots learn to dynamically adapt to the evolving constraints of long-term projects?

To explore these questions, interviews and surveys with developers could identify common customization needs across various projects. Scenario-based testing, where developers adjust bot settings and observe effects on project outcomes, could reveal which constraints are essential versus optional. Reinforcement learning techniques could be tested to develop adaptive bots that adapt to changing project constraints based on developer feedback.

\subsection{Human Moderation and Intervention Strategies in Multi-Bot Systems}

As multi-bot systems become more common, developers will need strategies to manage and intervene in these workflows. Determining the right level of control for developers in bot-driven systems is essential to maintaining productivity and flexibility.

\textbf{Research Questions:} What strategies can developers use to effectively moderate and intervene in multi-bot systems? What level of control should developers have over bots’ decisions and actions? How can systems be designed to allow for easy prioritization and intervention without disrupting the autonomous workflow of bots?

Data from user studies in simulated multi-bot environments could reveal how developers prioritize and intervene in bot-driven workflows. Task-switching scenarios can shed light on effective intervention strategies and prioritization needs. Prototyping different levels of control and tracking developer satisfaction and productivity in these configurations could inform optimal control schemes for bot-driven systems.

\subsection{Developing Metrics for Autonomous Bot Performance and Reliability}

For bots to be reliable team members, we need metrics that accurately measure their performance and reliability in different phases of development. These metrics will be essential for evaluating the effectiveness of autonomous bots and ensuring that they meet project goals~\cite{hernandez2017evaluation}.

\textbf{Research Questions:} How can the performance and reliability of bots be accurately measured in a bot-driven development environment? What metrics are most relevant for evaluating the effectiveness of autonomous bots in different stages of development? How can real-time metrics support decision making for developers who oversee bot-driven systems?

Experiments could test metrics such as code quality, bug detection accuracy, task completion time, and dependency updates. Bot-generated outputs could be tracked over time and compared to human-generated outputs, creating a benchmark for performance. Machine learning could be applied to develop predictive metrics that identify when bots are likely to encounter issues, allowing preemptive human intervention.

\subsection{Ethical, Legal, and Sustainability Considerations}

As bots assume more responsibility in their development, ethical and legal challenges arise. Ensuring accountability and ethical practices in bot-driven development will require new governance structures and clear guidelines~\cite{johnson2024role}.
Furthermore, the environmental impact and carbon footprint of these bots must be analyzed and minimized~\cite{shi2024greening}.

\textbf{Research Questions:} What are the ethical and legal considerations when bots generate code autonomously? How can accountability be ensured in cases where bots make independent decisions that lead to critical bugs or security issues? What guidelines and governance structures are needed to support ethical bot-driven development? Can we optimize these bots to minimize energy consumption while still remaining effective at their tasks?

Literature reviews and expert interviews in fields such as AI ethics and law could help identify emerging ethical concerns. Case studies in bot-driven environments would provide real-world examples of accountability and liability issues. Cross-disciplinary workshops with developers, ethicists, and legal experts could explore and refine best practices for governance in bot-driven development. Model optimization and configuration tuning techniques could help in making the underlying models of bots more sustainable.

\section{Concluding Remarks}

Bot-driven development (BotDD) presents an opportunity to rethink software engineering and pair programming by transforming bots from supportive tools into active, autonomous agents that manage and enhance development workflows. This shift brings both potential and complexity: as bots assume more responsibilities, new challenges emerge in skill acquisition, human-bot collaboration, and trust. Our research agenda outlines key areas for future investigation, with the aim of answering critical questions about bot customization, workflow integration, interruption tuning, and ethical governance. By addressing these questions, the field can better understand how to integrate bot-driven development effectively, supporting both faster development cycles and more adaptive workflows. Ultimately, bot-driven development could empower developers to focus on high-level orchestration and strategic oversight, creating a collaborative, resilient development environment where bots and humans work seamlessly together. As this paradigm evolves, it holds the potential to reshape software engineering practices, making them more efficient and aligned with the complexities of modern development demands.


\end{sloppy}
\end{document}